\documentclass[prd,twocolumn,amsmath,amssymb]{revtex4}
\usepackage{graphicx}
\usepackage{dcolumn}
\usepackage{bm}
\usepackage{color}
\usepackage{float}

\begin{document}
\title{Direct Measurements of Absolute Branching Fractions for $D^0$ and $D^+$ 
Inclusive Semimuonic Decays}

\author{
M.~Ablikim$^{1}$,              J.~Z.~Bai$^{1}$,   Y.~Bai$^{1}$,
Y.~Ban$^{11}$,
X.~Cai$^{1}$,                  H.~F.~Chen$^{15}$,
H.~S.~Chen$^{1}$,              H.~X.~Chen$^{1}$, 
J.~C.~Chen$^{1}$\footnote{Corresponding author\\ email address: 
chenjc@mail.ihep.ac.cn (J.C. Chen).},
Jin~Chen$^{1}$
% ...
\\
\vspace{0.2cm}
(BES Collaboration)}
\vspace{0.2cm} \affiliation{
\begin{minipage}{145mm}
$^{1}$ Institute of High Energy Physics, Beijing 100049, People's Republic of China\\
% ...
$^{11}$ Peking University, Beijing 100871, People's Republic of China\\
% ...
$^{15}$ University of Science and Technology of China, Hefei 230026,
People's Republic of China\\
% ...
\end{minipage}
}
\date{\today}
\begin{abstract}
By analyzing about 33 $\rm pb^{-1}$ data
sample collected at and around 3.773 GeV with the BES-II detector
at the BEPC collider, we directly measure the branching fractions for the neutral 
 and charged $D$ inclusive semimuonic decays to be
$BF(D^0 \to  \mu^+ X) =(6.8\pm 1.5\pm 0.7)\%$ and $BF(D^+ \to \mu^+ X) =(17.6 \pm 2.7 
\pm 1.8)\%$, and 
determine the ratio of the two branching fractions to be 
$\frac{BF( D^+ \to \mu^+ X)}{BF(D^0 \to  \mu^+ X)}=2.59\pm 0.70 \pm 0.25$. 
\end{abstract}
\maketitle

\section{Introduction}
The neutral and charged $D$ are both charmed mesons.
However, the lifetime of the $D^+$ meson
is surprised longer than that of the $D^0$ meson~\cite{pdg06}.
Isospin symmetry predicts that the partial widths to Cabibbo-favored 
semileptonic decays of the $D^0$ and $D^+$ mesons are equal. 
It is expected that the ratio of the branching fractions for the $D^0$ and $D^+$ 
inclusive semileptonic decays is
approximately equal to the ratio of $D^0$ and $D^+$ lifetimes 
up to the order of $O(\theta_c^2)$~\cite{pais}. 
Measurements of the branching fractions for the $D^0$ and $D^+$ semileptonic decays  
can provide valuable information on the difference in their lifetimes, which is important
to understand the charmed meson decays. 
Moreover, by comparing the branching fractions for the 
$D^0$ and $D^+$ inclusive decays with the sum of those for the 
exclusive decays,  one can estimate whether there are some decay
modes which are not observed yet.
 The $D$ mesons inclusive semielectronic decays
have been studied by several experiments~\cite{pdg06,cleoex,besex}.
However, the experimental studies for the $D$ mesons inclusive semimuonic decays 
 are very limited. Actually, there is no measurement 
 for the decay $D^+ \rightarrow \mu^+ X$ available 
($X$=any particles) in the PDG~\cite{pdg06} now.

In this Letter, we report direct measurements of the absolute branching fractions
for the inclusive decays $D^0\to \mu^+ X$ and 
$D^+\to \mu^+ X$, based on analyzing a data sample of about 33 pb$^{-1}$ 
collected at and around the center-of-mass energy ($\sqrt{s}$) 3.773 GeV with 
the BES-II detector. 

\section{The Beijing Spectrometer}
BESII is the upgraded version of the BES detector~\cite{BESII} operated at the 
Beijing Electron Positron Collider (BEPC)~\cite{BEPC}.
A 12-layer vertex chamber (VC)
surrounding the beam pipe provides trigger information.
A forty-layer main drift chamber (MDC) located outside the VC
performs trajectory and ionization energy loss ($dE/dx$) measurement
with a solid angle coverage of $85\%$ of $4\pi$ for charged tracks.
Momentum resolution of $\sigma_p/p = 1.7\%\sqrt{1+p^2}$ ($p$ in GeV/$c$)
and $dE/dx$ resolution of $8.5\%$ for Bhabha scattering electrons
are obtained for the data taken at $\sqrt{s}=3.773$ GeV. An array of 48 scintillation
counters surrounds the MDC and measures the time of flight (TOF) of charged
tracks with a resolution of about 200 ps for the electrons.
Surrounding the TOF is a 12-radiation-length, lead-gas barrel shower counter
 (BSC) operated in limited streamer mode, which measures the energies of electrons
and photons over $80\%$ of the total solid angle, and has an energy
resolution of $\sigma_E/E=0.22/\sqrt{E}$ ($E$ in GeV),
spatial resolutions of $\sigma_{\phi}=7.9$ mrad and $\sigma_Z=2.3$ cm for the
electrons. Outside of the BSC is a solenoidal magnet which provides a
0.4 T magnetic field in the central tracking region of the detector. Three
double-layer muon counters instrument the magnet flux return, and serve to
identify muon with momentum greater than 0.5 GeV/$c$. They cover $68\%$ of the
total solid angle with longitudinal (transverse) spatial resolution of 5 cm (3
cm). End-cap time-of-flight and shower counters extend coverage to the
forward and backward regions. A Monte Carlo  package based on GEANT3
has been developed for BESII detector simulation and 
comparisons with data show that the simulation is generally
satisfactory~\cite{simbes}.

\section{Data analysis} 
Around $3.773$ GeV, the
$\psi(3770)$ resonance is produced in electron-positron ($e^+e^-$) annihilation and
decays into $D\bar D$ pairs with a large branching fraction, 
$BF(\psi(3770)\rightarrow D\bar{D})=(85\pm 5)\%$~\cite{pdg06,nondd1,nondd2, 
nondd3,nondd4}.
If we can reconstruct $\bar D$ (they are
named as singly tagged $\bar D$) from the $D\bar D$ pairs, the other $D$ must 
exist
on the recoil side of the tagged $\bar D$. Taking this advantage, we can directly
measure the absolute branching fractions for the $D$ decays with the singly tagged 
$\bar {D}$ samples. 
The singly tagged $\bar D$ samples used in the analysis were reconstructed 
from the hadronic decay modes of $mKn\pi$ ($m$=0,1, 2; $n$=0, 1, 2, 3, 4)
in the previous works~\cite{597,608}. These give the total numbers of 
$7584\pm198{\rm (stat.)}\pm341{\rm (sys.)}$ singly tagged $\bar D^0$~\cite{597} and 
$5321\pm149{\rm (stat.)}\pm160{\rm (sys.)}$ singly tagged $D^-$~\cite{608}.  
Throughout the Letter, charge conjugation is implied. 

\subsection{Candidates for $D\rightarrow \mu^+ X$}
The candidates for $D\to \mu^+X$ are selected
in the system recoiling against the singly tagged $\bar D$ mesons.
It is required that the  candidate tracks should be well reconstructed in the MDC 
with good helix fits, and
satisfy $\rm |cos\theta |<0.67$, where $\theta$ is the polar angle.
Each track must originate from the interaction region, which requires
that the closest approach to the interaction point in the
$xy$-plane is less than 2.0 cm and in the $z$ direction is less than
20.0 cm. To reject the muons from kaon and pion decays,
the selected candidate tracks are required to originate from
the same vertex as those decay from the singly tagged $\bar 
D$ mesons, by requiring $\delta z < 2\sigma_z$ (2.0 cm), where
$\delta z$ is the minimum distance in the $z$ direction between the
candidate track and those decay from the singly tagged $\bar D$,  $\sigma_z$ is the 
standard deviation of the $\delta z$ distribution. To ensure 
that the track can be detected by the muon 
counter, the transverse momentum of each track is required to be 
greater than 0.52 GeV/$c$. Since the hit depth of muon in the
muon counter varies with its transverse momentum, the muon is selected 
by requiring that the track should hit at least one layer for the track with 
the transverse momentum in the region (0.52, 0.75) GeV/$c$,  two 
layers for the one in the region (0.75, 0.95) GeV/$c$, or three layers for the 
track with transverse momentum being greater than 0.95 GeV/$c$, respectively.
Because there is no charge symmetric background as those in the 
measurements of the semielectronic decay~\cite{cleoex,besex}, 
only "right-sign" muon candidates (with their charge opposite to the flavor 
of the single tag) are selected. 

\begin{figure}[thb]
\begin{center}
\includegraphics[height=9cm,width=9.cm]{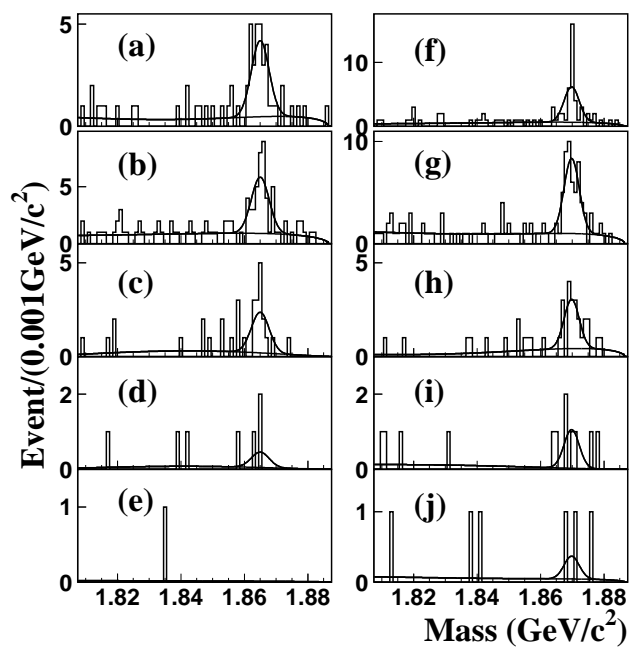}
\caption {Invariant mass spectra of the $mKn\pi$ combinations for the
events in which the muon candidates 
are observed on the recoil side against the $mKn\pi$ combinations   
for the singly tagged $\bar D^0$ (left column) and  $D^-$ (right column) 
mesons, where the tracks are with the transverse momenta in the region: 
(a) and (f) (0.52, 0.62) GeV/$c$;
(b) and (g) (0.62, 0.72) GeV/$c$;
(c) and (h) (0.72, 0.82) GeV/$c$;
(d) and (i) (0.82, 0.92) GeV/$c$ and
(e) and (j) (0.92, 1.02) GeV/$c$ intervals, respectively.}
\label{mass}
\end{center}
\end{figure}

Fig.~\ref{mass} shows the resulting invariant mass spectra of the $mKn\pi$
combinations for the events in which the candidates for muon are
observed on the recoil side of the $mKn\pi$ combinations for the singly tagged 
$\bar D^0$ (left column) and  $D^-$ (right column) mesons. 
Fitting each invariant mass spectrum with a Gaussian function for $\bar D$
signal and a special function~\cite{597} to describe the background shape,
we obtain number $N_{\rm obs}^{\mu}$ of the observed muon candidates. 
The total numbers of the observed 
candidates for $D^0\rightarrow \mu^+ X$ and $D^+\rightarrow \mu^+ X$ are
$79.3\pm 10.3$ and $99.6\pm 11.7$, respectively.

\subsection{Unfolding Procedure}
Since the muon, electron, kaon and pion may be misidentified to each other, 
the observed muon candidates consist of the true muons and those are misidentified 
from electron, kaon and pion. The true yield $N_{\rm true}^{\mu}$ of muons
can be extracted through an unfolding procedure. A detailed description
of the unfolding procedure can be found in Ref.~\cite{besex}. The right-sign
candidates are unfolded using the matrix Eq.~\ref{matrix}, 
\begin{equation} 
\begin{pmatrix}
N_{\rm obs}^{\mu} \\
N_{\rm obs}^{e} \\
N_{\rm obs}^{K} \\
N_{\rm obs}^{\pi} \\
\end{pmatrix}
= 
\begin{pmatrix}
\eta^{\mu\to \mu} & \eta^{e\to\mu} &\eta^{K\to\mu}&\eta^{\pi\to\mu} \\
\eta^{\mu\to e} & \eta^{e\to e} &\eta^{K\to e} &\eta^{\pi\to e} \\
\eta^{\mu\to K} & \eta^{e\to K} &\eta^{K\to K} &\eta^{\pi\to K} \\
\eta^{\mu\to \pi} & \eta^{e\to \pi} &\eta^{K\to \pi} &\eta^{\pi\to\pi} \\
\end{pmatrix}
\begin{pmatrix}
N_{\rm true}^{\mu} \\
N_{\rm true}^{e}   \\
N_{\rm true}^{K} \\
N_{\rm true}^{\pi} \\
\end{pmatrix},
\label{matrix}
\end{equation}
where $N_{\rm obs}^{a}$ and $N_{\rm true}^{a}$ ($a$ and $b$ denote muon, electron, 
kaon or pion) are the number of the observed candidates for $a$ and 
the number of the true $a$ particles, respectively; $\eta^{b \to a}$ is the rate of 
misidentifying the particle $b$ as $a$ ($a\neq b$) or the 
efficiency of identifying the particle $a$ ($a=b$). 

To remove the misidentified particles from the
observed muon candidate sample, we need to know
the numbers $N^{e}_{\rm obs}$, $N^{K}_{\rm obs}$ and $N^{\pi}_{\rm obs}$
of electrons, kaons and pions besides the number $N_{\rm obs}^{\mu}$ of muons.
They are also selected in the system recoiling against the singly tagged
$\bar D$ mesons. 
Electron, kaon and pion are identified by using the $dE/dx$, TOF and BSC
measurements, with which the combined confidence levels for the electron,
kaon and pion hypotheses ($CL_{e}$, $CL_{K}$ and $CL_{\pi}$) are calculated.
An electron candidate is required to have $CL_{e}>0.1\%$ and
$CL_{e}/(CL_{e}+CL_{K}+CL_{\pi})>0.8$, and a kaon candidate
is required to satisfy $CL_{K}>CL_{\pi}$ and $CL_{K}>0.1\%$.
The tracks not satisfying the selection criteria of muon, electron and kaon 
are treated as pions.

\begin{figure}[thb]
\begin{center}
\includegraphics[height=10cm,width=10cm]{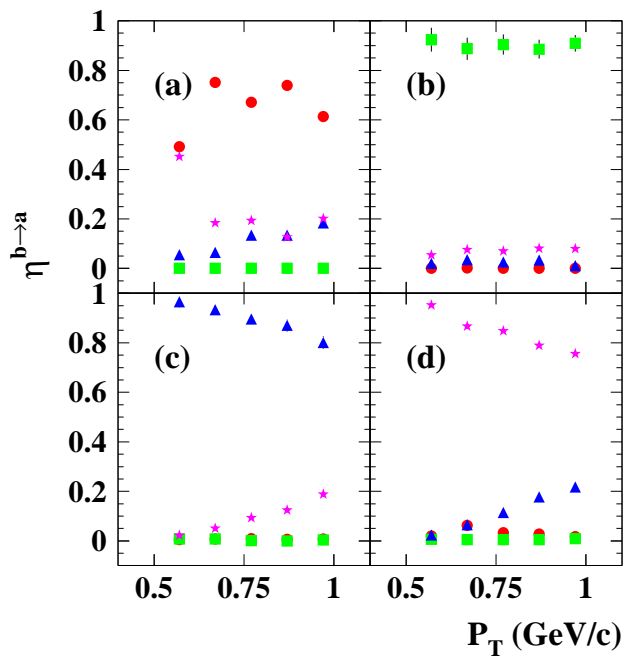}
\caption {$\eta^{b \to a}$ ($a$ and $b$ denote muon, electron, kaon or pion)
are the rates of misidentifying the particle $b$ as $a$ ($a\neq b$) or the 
efficiency of identifying the particle $a$ ($a=b$), where the dots, squares, 
triangles and stars are for the particles (mis)identified as
muon, electron, kaon and pion from the pure particle samples:   
(a) the muon sample selected from the cosmic rays; (b) the electron sample; 
(c) the kaon sample and (d) the pion sample.}
\label{pidmu}
\end{center}
\end{figure}

To obtain the number of the true muons, we also need to estimate the rates
$\eta^{b \to a}$ by analyzing pure muon, electron, kaon 
and pion samples. The muon sample is selected from cosmic rays.
The electron sample is selected
from the radiative Bhabha events. The pion and kaon samples are
selected from the $J/\psi\to\omega\pi^+\pi^-$ and
$J/\psi\to\phi K^+K^-$ events, respectively. 
The difference between the $D\bar{D}$ decay event environment and the selected particle 
sample environment 
is studied with two Monte Carlo matrices. One matrix is determined using the 
muons. electrons, kaons and pions from the Monte Carlo simulated cosmic ray, Bhabha. 
$J/\psi\rightarrow \phi K^+K^-$ and $J/\psi\rightarrow \omega \pi^+ \pi^-$ decays, and
another matrix is produced using the particles from the $D\bar{D}$ Monte Carlo 
samples. The difference in the unfold yields with various matrices 
are about $3.9\%$ for $D^0$ decay and $2.1\%$ for $D^+$ decay.
Since the rates $\eta^{b \to a}$ vary with momenta of the particles, 
we divide the transverse momentum region  (0.52,1.02) GeV/$c$ into five 
intervals in the analysis. Figure \ref{pidmu} shows the rates
 $\eta^{b\to a}$ in each transverse 
momentum interval obtained from the pure particle samples.

Inserting the numbers of $N^{a,i}_{\rm obs}$, the rates $\eta^{b\to a,i}$
($i$ denotes the $i$th transverse momentum interval)
in the matrix Eq.~\ref{matrix},
we obtain the yield $N_{\rm true}^{\mu,i}$ of the true muons
in the $i$th momentum interval. Summing the $N_{\rm true}^{\mu,i}$ over
the intervals, gives the total numbers of the true yield 
candidates to be $87.5\pm17.1$ for $D^0 \to \mu^+X$ and $129.3\pm19.5$
for $D^+\to \mu^+X$, respectively.

\subsection{Other Background}
In the selected candidate events, there are still some
contaminations from decays $K^+\to \mu^+\nu_\mu$ and $\pi^+\to \mu^+\nu_\mu$.
These backgrounds are estimated based on the Monte Carlo simulation.
The simulated background events are generated as $e^+e^-\to D\bar D$
events, where $D$ and $\bar D$ mesons are set to decay into all
possible modes except $D^0 \to \mu^+X$ for studying the $D^0$ decay, or 
except $D^+ \to \mu^+X$ for studying the $D^+$ decay, respectively.
In the Monte Carlo simulation, 
the decay modes of $D$ mesons and their branching fractions are
quoted from PDG~\cite{pdg06}, and the particle trajectories are simulated with the
GEANT3 based Monte Carlo simulation package of the BES-II
detector~\cite{simbes}. The numbers of these background events are
estimated to be $8.4\pm1.8$ and $3.0\pm1.6$ for $D^0\to \mu^+X$
and $D^+\to \mu^+X$, respectively.
After subtracting these background events, we obtain $79.1\pm17.2$ and
$126.3\pm19.6$ signal events for $D^0\to \mu^+X$ and
$D^+\to \mu^+X$, respectively.

\section{Results}
The branching fraction for $D \to \mu^+ X$ is determined by 
\begin{equation}
BF(D \to \mu^+ X)=\frac{N_{D \to \mu^+ X}}{N_{\bar D}\times
\epsilon_{D \to \mu^+ X}}, 
\label{br2}
\end{equation}
where $N_{D \to \mu^+ X}$  is the number of the  
events for $D \to \mu^+ X$,  $N_{\bar D}$ is the number of the singly 
tagged $\bar D$ mesons, $\epsilon_{D \to \mu^+ X}$ is the detection 
efficiency for $D \to \mu^+ X$.

The detection efficiency $\epsilon_{D \to \mu^+ X}$ 
is estimated with the Monte Carlo simulation. The
Monte Carlo events are generated as $e^+e^-\to D\bar D$, 
where $\bar D$ decay into the singly tagged $\bar D$ modes and
$D$ decay into semimuonic modes. The semileptonic decays are generated with 
the $q^2$ dependence of form factors given by the pole model~\cite{pole}. 
The generator has been applied in the previous 
measurements and reproduces exclusive semileptonic spectra 
well~\cite{dpk0muv,d0kmuv,dpkev,d0kev,dkpiev}.
In order to study the model dependence of the efficiency,
the semileptonic decays are also simulated with
the ISGW2 form factor model~\cite{isgw2}. The difference in
efficiencies determined with various form factor models is about $5.8\%$.
The averaged efficiency is determined by weighting
the branching fractions of the $D$ meson semimuonic 
decays~\cite{pdg06} and the numbers of the singly tagged $\bar{D}$ events. 
The efficiencies are $(15.4\pm0.2)\%$ for $D^0 \to \mu^+ X$ and $(13.5\pm0.2)\%$ for 
$D^+ \to \mu^+ X$, respectively.

Inserting the number $N_{D \to \mu^+ X}$ of the signal events for $D \to \mu^+ X$,
the number $N_{\bar D}$ of the singly tagged $\bar D$ mesons and the detection
efficiency $\epsilon_{D \to \mu^+ X}$ in Eq. (\ref{br2}), we obtain the branching
fractions for $D \to \mu^+ X$ to be $$BF(D^0 \to \mu^+ X)=(6.8\pm1.5\pm0.7)\%$$
and $$BF(D^+ \to \mu^+ X)=(17.6\pm2.7\pm1.8)\%,$$ where the first errors are
statistical and the second systematic. 

The systematic errors arise mainly from the
uncertainties in particle identification ($\sim 5 \%$ for muon~\cite{610}), in
tracking ($\sim 2.0\%$ per track), in the number of the singly tagged $\bar D$
mesons ($\sim 4.5\%$ for $\bar D^0$~\cite{597} and $\sim 3.0\%$ for
$D^-$~\cite{608}), in the $\delta z$ selection criterion ($\sim 3.5 \%$), in 
the input form factor models ($\sim 5.8\%$), in the 
 selected sample environment ($\sim 3.9\%$ for $D^0$ 
and $2.1\%$ for $D^+$), and in
the Monte Carlo sample statistics ($\sim 1.5 \%$). The systematic errors from the  
uncertainties of the $\eta^{b \to a}$ in the unfolding procedure are 
estimated using the Monte Carlo samples generated with the Gaussian 
distribution to describe the uncertainties of the $\eta^{b \to a}$ matrices,  
and they are estimated to be $\sim 2.1 \%$ for $D^0$ and 
$\sim 1.0 \%$ for $D^+$, respectively. 
The systematic errors from the poorly measured exclusive 
semileptonic decay modes are estimated using $D\bar D$ Monte Carlo 
samples generated with or without those modes, they are about $3.3\%$ and 
$4.7\%$ for $D^0\to \mu^+ \nu_{\mu}$ and $D^+\to \mu^+ \nu_{\mu}$, respectively.
Adding these uncertainties in
quadrature yields the total systematic errors to be 11.0\% and 10.3\% for $D^0 \to
\mu^+ X$ and $D^+ \to \mu^+ X$, respectively.

With the measured branching fractions for $D^0 \to \mu^+ X$ and
$D^+ \to \mu^+ X$, the ratio of the two branching fractions 
is determined to be
$$\frac{BF(D^+ \to \mu^+ X)}{BF(D^0 \to \mu^+ X)}=2.59\pm0.70\pm0.25,$$ 
where the first error is statistical and the second systematic 
arising from the uncertainties in the number of the singly tagged
$\bar D$ mesons, in the $\eta^{b \to a}$, in the Monte Carlo sample statistics, in the
 the selected sample environment and in the poorly measured 
decay modes. 

Table \ref{bijiao} shows the comparisons of the measured
branching fractions for $D \to \mu^+ X$ by the BES Collaboration with 
those measured by the ARGUS~\cite{argus}, CHORUS \cite{chorus} Collaborations and 
the averaged value from PDG~\cite{pdg06}. The measured $BF(D^0 \to \mu^+ X)$ 
is in good agreement with the measurements from other Collaborations.
\begin{table*}[htbp]
\caption{Comparisons of the measured branching fractions
for the inclusive semimuonic decays of $D$ mesons with those
measured by ARGUS~\cite{argus},
CHORUS~\cite{chorus} Collaborations and the PDG values~\cite{pdg06}.}
\label{bijiao}
\begin{center}
\begin{tabular}{ccccc} \hline \hline
 &BES & ARGUS&CHORUS&PDG06 \\ \hline
$BF(D^0\to \mu^+ X)$[\%]
&6.8$\pm$1.5$\pm$0.7&6.0$\pm$0.7$\pm$1.2&6.5$\pm$1.2$\pm$0.3 &6.5$\pm$0.7 \\
$BF(D^+\to \mu^+ X)$[\%]&17.6$\pm$2.7$\pm$1.8&-&-&- \\
$\frac{BF(D^+\to \mu^+ X)}{BF(D^0\to \mu^+
X)}$&2.59$\pm$0.70$\pm$0.25&-&-&- \\
$\frac{\tau_{D^+}}{\tau_{D^0}}$&-&-&-&2.54$\pm$0.02 \\
\hline \hline
\end{tabular}
\end{center}
\end{table*}

\section{Summary}
Using the data sample of about 33 $\rm pb^{-1}$
collected at and around $\sqrt{s}=3.773$ GeV with the BES-II
detector at the BEPC collider, we have studied the inclusive semimuonic
decays of $D$ mesons. The absolute branching fractions for $D^0 \to \mu^+ X$
and $D^+\to \mu^+ X$ are measured to be
$BF(D^0 \to \mu^+ X)=(6.8\pm 1.5\pm0.7)\%$ and $BF(D^+
\to \mu^+ X)=(17.6\pm 2.7\pm 1.8)\%$. The latter one 
is the first measurement. 
With the measured branching fractions for $D^0 \to \mu^+ X$ and
$D^+ \to \mu^+ X$, the ratio of the two branching fractions is
determined to be 
$BF(D^+\to \mu^+ X)/BF(D^0 \to \mu^+ X)=2.59\pm 0.70\pm0.25$,
which is consistent with the ratio of the lifetimes of $D^+$ and $D^0$
mesons, $\tau_{D^+}/\tau_{D^0}=2.54\pm0.02$~\cite{pdg06}. 

\section{Acknowledgments}
The BES collaboration thanks the staff of BEPC and computing
center for their hard efforts. This work is supported in part by
the National Natural Science Foundation of China under contracts
Nos. 10491300, 10225524, 10225525, 10425523, 10625524, 10521003, the Chinese Academy
of Sciences under contract No. KJ 95T-03, the 100 Talents Program
of CAS under Contract Nos. U-11, U-24, U-25, and the Knowledge
Innovation Project of CAS under Contract Nos. U-602, U-34 (IHEP),
and the National Natural Science Foundation of China under
Contract No. 10225522 (Tsinghua University).

\end{document}